\documentclass[12pt]{article}
\usepackage{graphicx} 	
\usepackage{amsmath} 	
\usepackage{array} 	
\usepackage{hhline}
\usepackage{pslatex}    
\usepackage{theorem} 

\theoremheaderfont{\scshape}
\theoremstyle{plain}
\theorembodyfont{\upshape}

\author{Giovanni Gasparri\\
  2ware Srl - Italy\\
  \makeatletter
  \texttt{gasparri@2ware.it}
  \makeatother
}
\title{\textbf{Using image partitions in $4^{th}$ Dimension}\footnote{ACI, 4D, $4^{th}$ Dimension - Copyright  pending. All rights reserved.}}
\date{
  {\small October $26^{th}$, 2004}}
\begin{document}
\markright{Using image partitions in $4^{th}$ Dimension - Giovanni Gasparri}
\maketitle
\pagestyle{myheadings}

\begin{figure}
\label{F1} \centering
\includegraphics[width=0.3\textwidth]{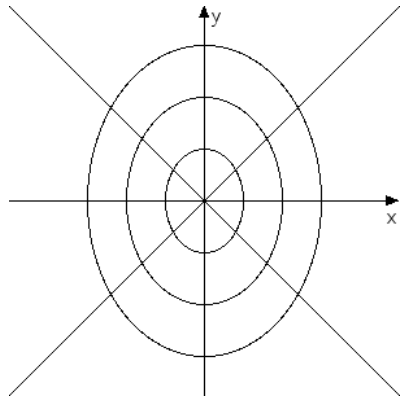}
\caption{Basic image}
\end{figure}

\textbf{
I have plotted an image (Figure \ref{F1}) by using mathematical functions. I'Õm going to show an alternative method to: detect which sector has been clicked; highlight it and combine it with other sectors already highlighted; store the graph information in an efficient way; load and splat image layers to reconstruct the stored graph.
}

\section{Detecting the clicked area}

Instead of using the button grid method or the generalized image maps \cite{A2}, to discretize the space and approximate the curves with rectangles, IÕ'm using an alternative system to detect the clicked image partition based on a mathematical continuous representation.
Any window can be considered as a cartesian plane having the upper/left corner as center and having the $y$ axes reversed. One can make a translation so that the center of the window coincides with the center of the cartesian plane as well.

Let $(x_0,y_0)$ be the new center;
let $(m_x,m_y)$ be the clicked point; the following procedure shows how to evaluate the translated coordinates $(x,y)$.

\begin{verbatim}
  GET WINDOW RECT(wleft;wtop;wrigth;wbottom)
  GET MOUSE(MX;MY;button;*)
  MX:=MX-(wleft)
  MY:=MY-(wtop)
  x:=MX-X0
  y:=Y0-MY
\end{verbatim}

At first one has to evaluate the angle $\theta=\arctan\frac{y}{x}$ and then adjust it (by properly adding multiples of $\pi$  in order to single the correct quadrant out) to get  a value between $0$ and $2 \pi$.\\

The following step is to detect in which ellipse the clicked point is included.
Let $b_i, a_i, i=1,\cdots,3$ be the $x-$ and $y-$ semi axis of the ellipse $i$.
If $\frac{x^{2}}{i^{2}b^{2}}+\frac{y^{2}}{i^{2}a^{2}}<=1$ is satisfied then the point $(m_x,m_y)$ belongs to the interior of the ellipse $i$.\\
Since the ellipses are concentrical, if a point belongs to the smallest one it belongs to the others as well.
A case statement can easily manage this situation.

\section{Highlighting method}
\begin{figure}
\label{F2} \clearpage \centering
\includegraphics[width=0.5\textwidth]{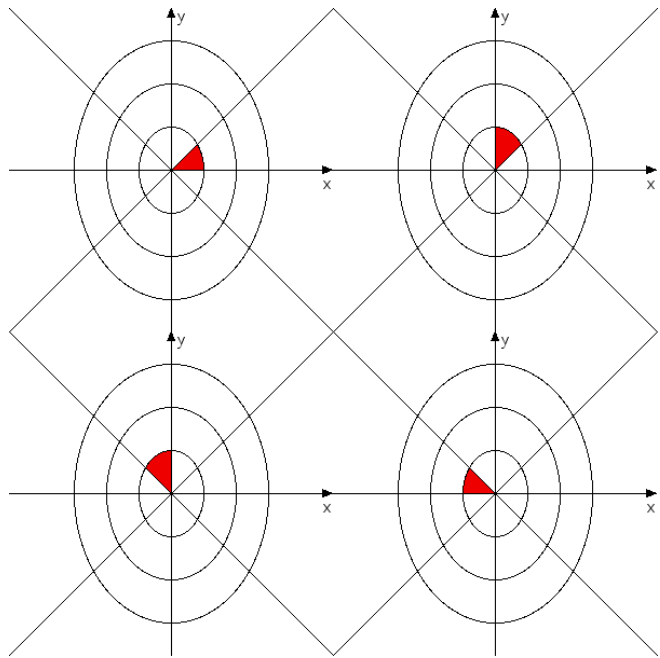}
\caption{Image Sequence form Picture Library}
\end{figure}

I have put the basic image (Figure \ref{F1}) in the Picture Library. I have created 24 sequential pictures like those one shown in Figure \ref{F2}, each of them representing a colored sector of the image. They fill the resource range from 2025 \textit{(LibraryStart)} to 2049.\\

When a sector is chosen the corresponding image is loaded from the library and using the Exclusive Superimposition Picture operator (XOR), the new image is built.

\begin{verbatim}
  GET PICTURE FROM LIBRARY(LibraryStart + selected;$rif)
  vpimage:=$skeletron | (vpimage & $rif)
\end{verbatim}

\section{Efficient Storage}
\begin{figure}
\label{F3} \centering
\includegraphics[width=0.5\textwidth]{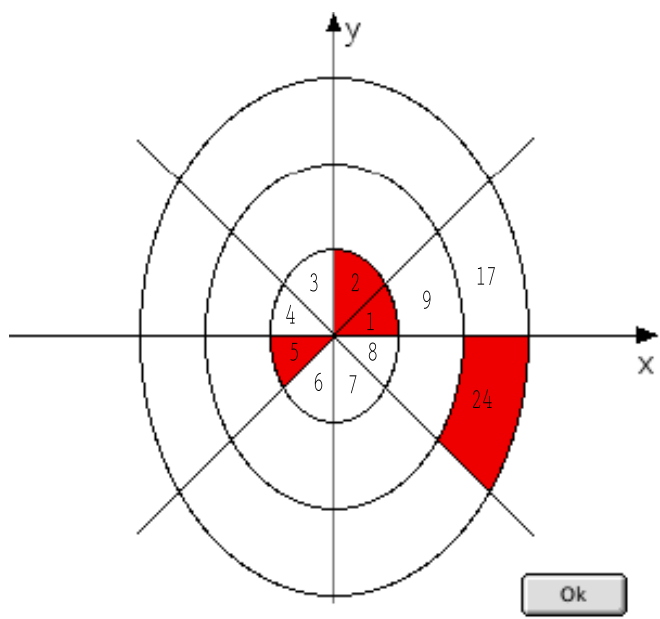}
\caption{Graph representing the bit sequence 100000000000000000010011}
\end{figure}

Whenever a partition is highlighted a longint variable called \verb"status" is updated using bitwise operators.\\
The variable is used in binary mode. The last ordered sequence of 24 bits records the status of image sectors. The rightmost bit is therefore the first. For example, a bit sequence as ``\textbf{100000000000000000010011}'' implies that the first, the second, the $5^{th}$ and the last one are highlighted (Figure \ref{F3}).

The code below shows how to update the variable status.

\begin{verbatim}
  C_LONGINT($temp)
  $temp:=0 ?+ selected
  status:=$ status  ^| $temp
\end{verbatim}

Since the graph information fill just 24 bits it represents the optimal storage way in terms of memory.
No pictures, blobs or similar objects are stored in the database. A \textit{longint} field allows queries on it even if this is a very time consuming operation.\\
\\
When an existing record is loaded then a new image is built by reading its own bit sequence and applying the Inclusive Superimposition Picture operator (OR) for each ``\textbf{1}'' found. This procedure gets the corresponding picture from the library and superimposes it on the existing image of the screen .

\begin{verbatim}
  For ($i;1;24)
    If (status ?? $i)
      GET PICTURE FROM LIBRARY(LibraryStart + $i;$vptemp)
      vpImage:=vpImage | $vptemp
    End if 
  End for
\end{verbatim}

\section{Conclusions}
The detecting method presented is useful whenever one has to superimpose an hypothetical mask layer to an image, the mask being represented by mathematical objects having the same center, such as concentrical ellipses or circumferences. This method allows to manage easily the detecting process using a single procedure.\\
The proposed highlighting method is highly efficient whenever one has to handle a limited number of sectors and the images have few colors or can be drawn using vectorial graphic software. The storage method allows the database to hold just the information concerning the highlighted sectors, without keeping track of the specific coordinates.  
\bibliographystyle{plain}\bibliography{/path/to/your/bib/file/here}

\begin{thebibliography}{}
\bibitem{A1} Jean-Yves Fock-Hoon,  Using the Picture Library in 4D,  Tecnical Note 02-39 (2002).
\bibitem{A2} Jeremy Sullivan,  Image Maps,  Tecnical Note 02-44 (2002).
\end{thebibliography}

\end{document}